\documentclass[prl,twocolumn,groupedaddress]{revtex4}

\usepackage[dvips]{graphicx}

\begin{document}

\newcommand{\av}[1]{\left\langle#1\right\rangle}
\newcommand{\£}{\pounds}
\newcommand{\E}{\mathrm{E}}
\newcommand{\Var}{\mathrm{Var}}
\newcommand{\Cov}{\mathrm{Cov}}
\newcommand{\x}{\underline{x}}
\newcommand{\C}{{\bf C}}
\newcommand{\D}{{\bf D}}

\title{Impact of Unexpected Events, Shocking News  and Rumours\\
on Foreign Exchange Market Dynamics}
\author{Mark McDonald$^{1}$, Omer Suleman$^2$, Stacy Williams$^3$, Sam Howison$^1$ 
and Neil F. Johnson$^2$}
\address{$^1$Mathematics Department, Oxford University, Oxford, OX1 2EL, U.K. \\
$^2$Physics Department, Oxford University, Oxford OX1 3PU, U.K.  \\
$^3$FX Research and Trading Group, HSBC Bank, 8 Canada Square, London E14 5HQ, 
U.K.}

\date{\today}

\begin{abstract}
We analyze the dynamical response of the world's financial community to various types of unexpected events, including the 9/11 terrorist attacks  as they unfolded on a minute-by-minute basis. We find that there are various `species' of news, characterized by  how quickly the news gets absorbed, how much meaning and importance is assigned to it by the community, and what subsequent actions are then taken. For example, the response to the unfolding events of 9/11 shows a gradual collective understanding of what was happening, rather than an immediate realization. For news items which are not simple economic statements, and hence whose implications are not immediately obvious, we uncover periods of collective discovery during which collective opinions seem to oscillate in a remarkably synchronized way. In the case of a rumour, our findings also provide a concrete example of  contagion in inter-connected communities. Practical applications of this work include the possibility of producing selective newsfeeds for specific communities, based on their likely impact.

\end{abstract}

\maketitle

\section{I. Introduction}
Governments and planning agencies often second-guess people's collective response to important pieces of news and rumours -- for example, stories concerning terrorist attacks, security risks, health scares and economic upsets. Timing and content are everything, hence the often-cited claim that  governments tend to bury bad news on a bad news day \cite{bbc}. 

In a commercial setting, the fact that the amount of information available on the web is becoming overwhelming implies that there are significant opportunities for any content provider who can provide a customized stream of important news items. Of course, the notion of `importance' is somewhat personal -- it depends on how surprising you find such a piece of news, and how relevant it seems to your affairs. 

From a physics perspective, any piece of news represents a perturbation to the many-body financial system, which itself comprises a complex global web of financial traders. The typical approach of a condensed matter physicist attempting to understand the properties of such a collective or `complex' many-body system, is to measure the system's response to such a perturbation in order to build up a picture of how the system is wired. In other words, a good way of understanding what is in such a closed box is to shake it. But can such an approach work for human-based Complex Systems? This paper attempts to take a first step along this path, by monitoring the responses of the global foreign exchange (FX) market to a variety of perturbations.

Financial markets are continually being `kicked' by news and events. Recent research suggests that most everyday news items have relatively little effect on market movements \cite{olsen}. But what about major news? Unlike most physical systems such as a gas of electrons, individual humans may not all respond to external shocks in the same way. Worse still, even if a given piece of news is announced globally at a given moment, it will tend to reach different people at different times. Furthermore their response times will tend to differ, as will the extent to which they believe the news to be true. As if this wasn't complicated enough, the way in which people respond will also differ in general. What is good news for one person or market, may be bad for another -- or just plain irrelevant. One might therefore wonder if any kind of systematic, coordinated, or synchronized response will ever be observed. This paper shows that indeed it can, but that the type of response depends strongly on the `species' of news.

The September 11 terrorist attacks in the U.S. were unambiguously bad in terms of human lives and world affairs. But how is such news to be interpreted from the point of view of a trader dealing with a particular currency?  If it is bad for the U.S. Dollar, is it therefore good for the Yen? After all, to sell a currency in the foreign exchange (FX) markets, you need to buy another one -- which is why it is called an exchange-rate. But in a moment of supposed global crisis, there is presumably no unambiguously safe currency to buy. So how should the unfolding of such news affect traders' decisions to buy or sell? Do they sell U.S. dollars and buy Japanese Yen, or vice versa -- or maybe do something entirely different? Or is there instead some more gradual process of collective learning -- a sort of collective realization or discovery of the meaning and implications of such news?

This paper attempts to address these issues using clustering methods \cite{MSWHJ2005} as explained in Sec. II. In Sec. III we describe the datasets themselves, before moving in Sec. IV to an analysis of the responses to the news. We find that similar results are seen for similar species of news. We focus on the following four case-studies: 

\noindent (1) The terrorist attacks of 11 September 2001 in the U.S.; 

\noindent (2) A false rumour that the Chinese currency (CNY) was about to be revalued (11 May 2005); 

\noindent (3) The real CNY revaluation (21 July 2005); 

\noindent (4) A particular piece of economic news in the form of an announced government statistic which turned out to be significantly different from that expected. 

\noindent In Section V we establish that these results are robust to the choice of particular clustering method. Section VI presents a more speculative discussion of the meaning of news, and the implications of our findings for the  classification of news into `species'. Section VII provides the concluding summary.

\section{II.  Cluster Analysis}

Given that we measure the dynamical response using cluster analysis of exchange-rate correlations, it is important to establish exactly what we mean by such cluster analysis. This is the discussion that we now undertake in this section.

Cluster analysis is a field of statistics which attempts to classify objects into groups. When performing a cluster analysis, one must first choose a measure of the proximity between different objects, the dissimilarity measure, and then specify an algorithm which groups objects into clusters, the cluster algorithm. One important set of clustering algorithms are agglomerative hierarchical cluster algorithms. Performing such analyses on a set of $N$ timeseries results in an indexed hierarchy, where each level of clustering is covered, beginning from $N$ clusters of one object each, and progressing until all the objects are in one cluster containing $N$ objects. The index details the distance at which two clusters are merged. 

A frequently used measure of the similarity of two timeseries is their correlation,
\begin{eqnarray}
\rho_{ij} & = & \frac{\E (X_i-\E(X_i))(X_j-\E(X_j)) }{\sqrt{\Var(X_i)\Var(X_j)}},
\end{eqnarray}
which is often estimated by Pearson's product-moment correlation coefficient
\begin{eqnarray}
\label{eqn:Pearson}
\hat{\rho}_{ij} & = & \frac{\sum_k(X_{ik}-\bar{X_{i}})(X_{jk}-\bar{X_j})}{\sqrt{\sum_k{(X_{ik}-\bar{X_i})^2} \sum_l{(X_{jl}-\bar{X_j})^2}}},
\end{eqnarray}
where $\bar{X_i}$ is the sample mean for $X_i$ and similarly for $\bar{X_j}$ and $X_j$. For a set of $N$ timeseries, one can form a correlation matrix $\C$, whose elements are $\C_{ij} = \hat{\rho}_{ij}$.

In Ref. \cite{MRN1999}, Mantegna proposed using Minimum Spanning Trees (MST) as a clustering procedure to identify hierarchical structure in financial markets. The procedure proposed was to transform the matrix $\C$ into a distance matrix, $\D$, where the elements of $\D$ are $d_{ij}=\sqrt{2(1-\hat{\rho}_{ij})}$. This distance matrix can be thought of as the adjacency matrix of a fully-connected, weighted $N$-graph, from which it is simple to construct the MST. Two algorithms commonly used to construct the MST are Kruskal's algorithm \cite{JBK1956} and Prim's algorithm \cite{RCP1957}. The MST has an indexed hierarchy associated with it, and it was this hierarchy that Mantegna suggested as a tool for defining a taxonomy of financial assets.

The dissimilarity measure chosen by Mantegna of $d_{ij}=\sqrt{2(1-\hat{\rho}_{ij})}$ is approximately the Standardized Euclidean distance between the timeseries $i$ and $j$ \cite{MS2000,KR1990}. Since there is evidence that the choice of dissimilarity measure has less effect on the clustering than the choice of clustering algorithm \cite{PH2003}, we continue to use this measure for comparison with previous work; however, we note that for the MST any co-monotonic transformation of distances would give the same cluster structure \cite{KR1990,SS1973}.

The first attempt to produce a hierarchy of financial securities was in Ref. \cite{BFK1966}; however, such techniques received little attention until Mantegna's paper \cite{MRN1999}. Since then, there has been an explosion of papers in the area from the field of econophysics \cite{MRN1999,BCLM2003,BCLM2004,OCKK2002,MBLM2003,OCKK2003a,OCKK2003b,OCKK2003c,OM2000,BVM2000,BLM2001,MAM2004,TLGM2005,MSWHJ2005}. The MST clustering algorithm has been applied to interest rate markets \cite{BGV2001, MAM2004}, equity market indices \cite{BVM2000} and Foreign Exchange (FX) markets \cite{MSWHJ2005}. In Ref.  \cite{TAMM2005}, the MST procedure was extended to include other links so that it was no longer a tree, but was the graph with the maximum amount of information which could be embedded in a surface. The graph which can be embedded in the surface of a sphere was termed the Planar Maximally Filtered Graph (PMFG). However, the hierarchical structure of this PMFG is identical to that of the MST and, as such, the procedure would not be of use here. The use of both the MST and the PMFG as tools for noise-undressing of correlation matrices has been investigated in Ref. \cite{TLGM2005}.

Using the MST as a clustering tool is the same procedure that statisticians refer to as the Single Linkage Clustering Algorithm (SLCA) \cite{GR1969}. This is one of the simplest hierarchical clustering algorithms, and has been in use as a clustering algorithm since 1951 \cite{FLPSZ1951}, although the first algorithm for calculating the MST was published in 1926 \cite{BJ1926}. However, the simplicity of the algorithm can sometimes be detrimental. When merging two clusters, the SLCA/MST method chooses the shortest link between the two clusters and then defines the distance between all objects in the two different clusters to be this smallest distance, before merging the two clusters into a new one, indexed by this distance. As the algorithm progresses, particularly for large $N$, there is a tendency for the method to link objects into chains, where the objects at each end of the chain have little or nothing in common with each other. For this reason it is usual to compare the results from one clustering method with other clustering methods. If several methods which group objects in very different ways agree on the cluster structure, then one can be confident of the robustness of ones results.

One cluster algorithm which groups objects in a very differrent way to the SLCA is the Complete Linkage Clustering Algorithm (CLCA). Where the SLCA defines the distance between two clusters as the minimum of the pairwise distances between two objects, one in each cluster, the CLCA defines the distance as the maximum pairwise distance. If two such radically different clustering algorithms agree on the cluster structure, this is very strong evidence of robustness. Whilst the SLCA groups objects into chains, the CLCA tends to group objects into compact groups, where unless all the objects in a group are very close together, they will not form a cluster until a very large distance. For this reason the SLCA is often termed as being \emph{space-contracting} and the CLCA as \emph{space-dilating} \cite{KR1990,SS1973}.

An algorithm which groups objects in a more moderate way is the Average Linkage Cluster Algorithm (ALCA). In this algorithm, the distance between two clusters being merged is the average of all the pairwise distances, one object in each cluster.

If the three algorithms above agree on the cluster structure, this is very strong evidence of the robustness of the structure being identified. {\em In this paper, we use all three to confirm the robustness of our results}. In addition, we use a further algorithm, Ward's Algorithm, which clusters objects by a minimizing variance procedure \cite{JHW1963}. If this method agrees with the three linkage methods, then this provides extremely strong evidence of robustness.

\begin{figure}[tp]
\begin{center}
\includegraphics[width=0.4\textwidth]{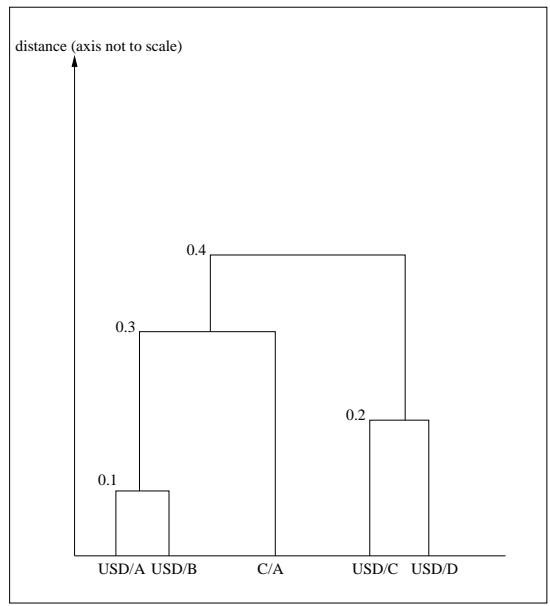}
\end{center}
\caption{(Color online) A schematic section of a hierarchy to illustrate the definition of the clustering distance for a currency. In this simple example, there are five currencies: A, B, C, D and USD (i.e. U.S. Dollar). The clustering distance for USD is 0.4, since one has to travel up the hierarchy to a distance of 0.4 until all the USD-based exchange rates are included in the same cluster.}
\label{fig:FXExample}
\end{figure}

All hierarchical cluster algorithms result in an indexed hierarchy. The distances of the objects in the fully connected $N$-graph reside in a Euclidean space, and so obey the metric inequality: $d_{ij} \leq d_{ik} + d_{kj}$. However, the correct space in which to describe a hierarchy is an \emph{ultrametric space} \cite{RTV1986,SS1973} in which the distances of objects under a clustering algorithm, $d_{ij}^{*}$, obey the stronger ultrametric inequality: $d_{ij}^{*} \leq \max(d_{ik}^{*},d_{kj}^{*})$. It is these ultrametric distances \cite{cophenetic} which we use to define the clustering distance for a currency. We explain the definition of this clustering distance with the schematic hierarchy shown in Figure \ref{fig:FXExample}.

Figure \ref{fig:FXExample} shows a section of a hypothetical hierarchy resulting from using a hierarchical clustering method on the exchange rates from five currencies: A, B, C, D and USD (i.e. U.S. Dollar). The clustering distance for a currency is the minimum distance at which one can partition the hierarchy and get all the base rates for that currency in the same cluster. Let us consider what the clustering distance is for USD for the case shown in Figure \ref{fig:FXExample}. In this example, if one partitions the hierarchy at 0.1, one has a cluster of (USD/A, USD/B), one cluster of (USD/C) and one cluster of (USD/D), so 0.1 is not the clustering distance. If one partitions the hierarchy at 0.2, there are two clusters containing USD based exchange rates: (USD/A,USD/B) and (USD/C,USD/D), so 0.2 cannot be the clustering distance. Partitioning the hierarchy at 0.3 still results in two clusters containing USD based exchange rates, one cluster of (USD/A,USD/B,C/A) and another cluster of (USD/C,USD/D). However, if one were to partition the hierarchy at 0.4, there would be one cluster containing (USD/A,USD/B,C/A,USD/C,USD/D). All the USD based exchange rates are in the same cluster---hence the clustering distance for USD in this rather simplistic example is 0.4. Note that as a result of the distance definition, a smaller distance corresponds to a stronger correlation.

\section{III. The Data}
The data used for this paper was provided by HSBC Bank. We use the last tick of each minute to form 1-minute prices for all exchange rates between the following currencies: Australian Dollar (AUD), Canadian Dollar (CAD), Swiss Franc (CHF), Euro (EUR), British Pound (GBP), Japanese Yen (JPY) and U.S. Dollar (USD) \cite{ISOCode}. This gives rise to 42 timeseries, from which we then form 1-minute log returns.

When performing regressions or calculating correlations, it is necessary to ensure that each timeseries in the data is \emph{autocovariance stationary} \cite{CA2001, GN1974}. To this end, Augmented Dickey-Fuller (ADF) tests (of various orders) \cite{DF1979, JGM1991, CA2001} were performed on the returns. The null hypothesis for this test is that the data is nonstationary. However, when ADF tests were performed on the return data, this null hypothesis is strongly rejected---evidence that the returns are autocovariance stationary.

Missing values were deleted prior to calculating correlations. To preserve the positive-definite property required of correlation matrices, if a timestep was deleted from one exchange rate, that timestep was deleted from all currency pairs under investigation.

\section{IV. Results}
We now turn to discuss each one of our four case-studies in detail. To investigate the dynamics of the clustering distance in each case, we use a window of length $T$=120 timesteps which we then move through the data one timestep at a time. There are $N$=42 timeseries, so $T \geq N$, as required. For each window the clustering distance for each currency was calculated, as described above. 
\begin{figure*}[tp]
\begin{center}
\includegraphics[width=0.95\textwidth]{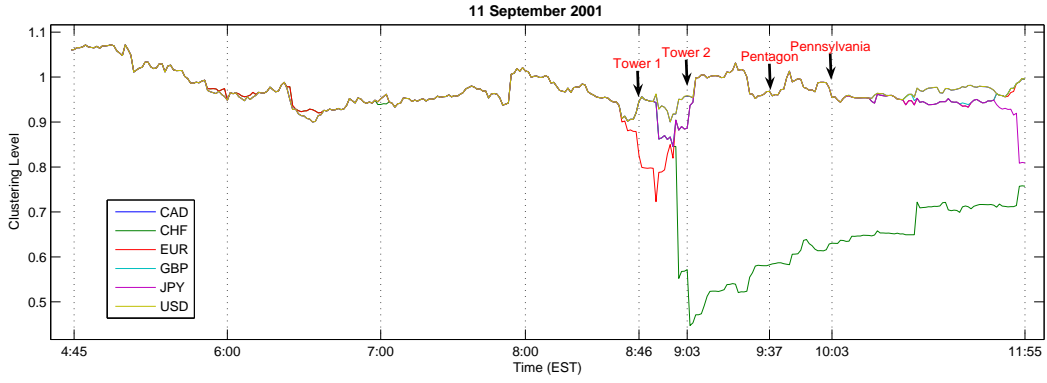}
\end{center}
\caption{(Color online) The clustering distances for September 11 2001.}
\label{fig:Sept11}
\end{figure*}
\subsection{1. 11 September, 2001}
The results for 11 September 2001,  are shown in Figure \ref{fig:Sept11}.
The equity markets closed as a result of the terrorist attacks, but the Foreign Exchange (FX) markets remained open thereby allowing us to investigate the effect of the unfolding news on global markets. Figure \ref{fig:Sept11} shows the power of this technique, with  the market impact by currency   clearly visible. 

The first feature to note in Figure \ref{fig:Sept11}, is a contagion-like effect whereby clustering gets successively transferred between currencies. Prior to the first tower being hit at 8:46am, there is no significant clustering in any currency, although the Euro seems to be slightly `in play' \cite{MSWHJ2005}. As soon as the first tower is hit  at 8:46am, a small amount of additional clustering is seen in the Euro. 

Remarkably, the Swiss Franc then undergoes a dramatic clustering {\em prior to the second tower being hit}. In other words, even though the second tower has not yet been hit -- and hence the hitting of the first tower can still be regarded as some form of awful accident -- traders have converged on the notion that something strange is happening. As a result, the Swiss Franc comes into play, taking centre stage in trading and hence giving rise to a very small clustering level as shown in Figure \ref{fig:Sept11}. At the same time, there seems to be a small split in opinion in that the Japanese Yen is very slightly in play as well. 
As soon as the second tower is hit, and people therefore realize that this is more than just two very unlikely accidents, the Swiss Franc undergoes another dramatic reduction in clustering level. In essence, activity in the Swiss Franc is then driving the market. 

Just as remarkable is the {\em absence} of any additional clustering features associated with the subsequent attacks, i.e. the Pentagon at 9:37am and the crash in Pennsylvania at 10:03am. It is as though the market had already decided that the news is as bad as it can get. Hence these two otherwise extremely dramatic events -- which on any other day would surely have driven the market into mayhem -- then had no visible effect. It is as if they hadn't happened.

\begin{figure}[tp]
\begin{center}
\includegraphics[width=0.45\textwidth]{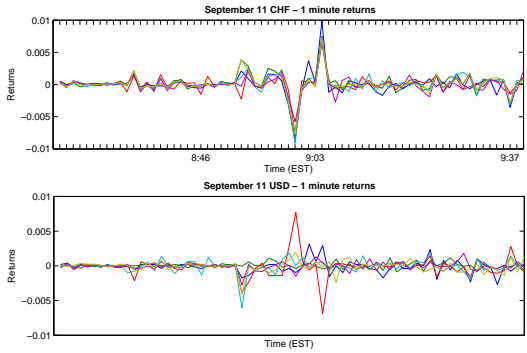}
\end{center}
\caption{(Color online) The returns for 11 September 2001, over a period of time which includes the first three attacks. Top panel: returns from the CHF-based exchange rates. Bottom panel: returns from the USD-based exchange rates.}
\label{fig:Sept11Returns}
\end{figure}

It is interesting that it is the Swiss Franc (CHF)  which is the clustered currency in Figure \ref{fig:Sept11} and not the U.S. Dollar (USD). One might think that the reason is trivial -- CHF may be acting as a `safe haven' currency which people buy in order to try to protect themselves in uncertain times. However, a quick look at the actual returns shows that this cannot be the complete answer.
In particular, Figure \ref{fig:Sept11Returns} shows the returns from a one-hour window for 11 September 2001. This one-hour period includes the first three attacks. The top panel shows the returns for the CHF-based exchange rates; the bottom panel the USD-based exchange rate returns over the same period. The horizontal time-axis is the same in both cases. It can be seen that before the first attack, the market is quiet. The first large return comes a short time after the first plane hits the World Trade Center. In other words, {\em the market takes time to absorb the news, assimilate its meaning and validity, interpret its importance, and hence react to it by trading}. It can be seen that the reactions of the CHF exchange-rates are effectively {\em synchronized}, and certainly more so than the USD exchange-rates. This is an important feature, since there is obviously no invisible hand or central controller coordinating such a collective dynamic -- and the fact that it is in the CHF-based rates but not the USD-based ones, implies that there is indeed a collective realization or `consciousness' that trading in CHF is the right thing to do.

Most importantly, it is not simply the case that people are buying CHF and selling everything else. To a first approximation, increased demand for CHF (i.e. increase number of buy orders) will increase its exchange-rate while reduced demand for CHF (i.e. increase number of sell orders) will decrease it. We can see from  Figure \ref{fig:Sept11Returns} that there is a spontaneous synchronization of buy {\em and} sell decisions which is occuring in oscillatory fashion -- in other words,  {\em synchronized} oscillatory fluctuations emerge from an otherwise uncoordinated system. For this reason, the rationale that people are simply seeking to buy CHF cannot be right. Instead, there appears to be a rich process of collective change-of-mind. Although one needs to be careful with such analogies, we believe that these dynamics represent one of the first ever demonstrations that a financial market comprising semi-autonomous traders, can show a collective conscience and/or concensus, and that this can then vary dynamically over time.

\begin{figure}[tp]
\begin{center}
\includegraphics[height=0.45\textwidth, angle=90]{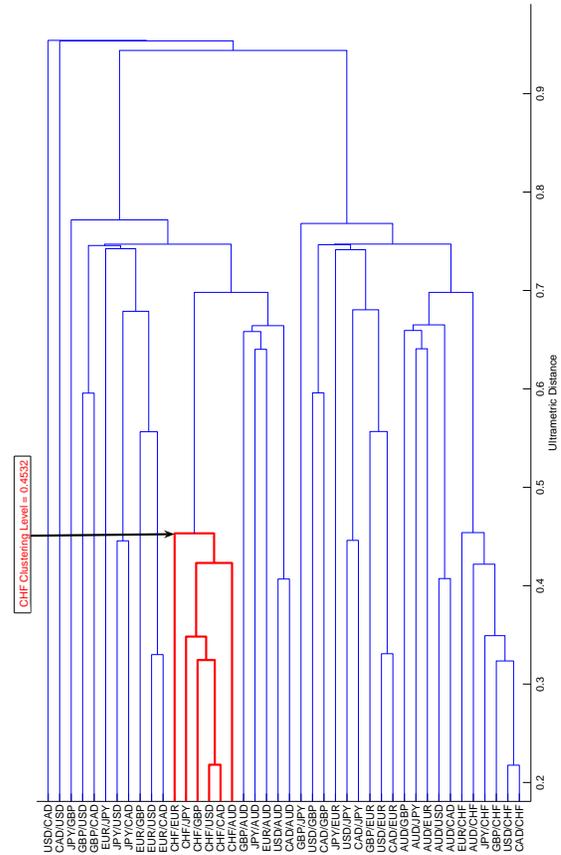}
\end{center}
\caption{(Color online) The hierarchy resulting from using the SLCA on a 120 timestep window ending at 9:05 EST on September 11 2001.}
\label{fig:dendrogramCHF}
\end{figure}

Even though this period of CHF-based synchronization is not long, the magnitude of the returns has a large effect on the correlation---hence the drop seen in the clustering distance. To illustrate precisely how clustered the CHF based exchange rates are at that time, Figure \ref{fig:dendrogramCHF} shows the hierarchy resulting from using the SLCA on a 120 timestep window of data ending at 9:05 EST on 11 September 2001. The CHF cluster is picked out in red \cite{inverseCluster}. It is evident that all the CHF based rates are clustered together. The clustering distance of 0.4532 corresponds to a correlation of 0.897.

\subsection{2. CNY Revaluation --- Rumour}
The 11 September analysis above shows that the SLCA clustering distance can identify a currency dominating a group of others in the market, even for 1-minute returns. In this section we consider a more market-specific shock. At 08:23 GMT, Wednesday 11 May 2005, a rumour began to propagate through the market that the Chinese government would imminently remove the peg between the Chinese currency (the Chinese Yuan Renminbi, CNY) and the USD \cite{peg}. It turned out that the rumour was false, but this was not known at the time. The credibility of the rumour, and hence its contagion period, survived for approximately 30 minutes. Toward the end of this period, it gradually became evident to most traders that the rumour was untrue. 

This example is of particular interest for two reasons. First, the shock is specific to the FX market, yet external to the currencies included in our analysis. Hence it does not represent news which is unambiguously `good' or `bad' for a given currency -- nor are the subsequent actions which a trader should take immediately obvious. Second, the real revaluation of the CNY happened only a couple of months later. Hence it is almost as if the same shock happened twice. Comparing the difference between the reaction of the market in both situations, therefore enables us to interpret how the market processes such information and to discover if the market `learns' in any way. In other words, does the FX market collectively go ahead and repeat its earlier behavior or do something completely different? A physical system would, in the absence of hysteresis effects, respond in a similar way if subject to similar conditions -- but it is not obvious whether human-based complex systems behave in such a way. This study therefore provides a remarkably unique experiment in a real-world system.

Prior to the rumour, the market was actually expecting some form of official announcement about revaluing the CNY in the ensuing days, weeks or months. In other words, the market was `susceptible' to such a rumour. The Chinese government had been under pressure for some time to revalue its currency, which other countries felt was fixed at an unrealistically low price. However, the knock-on effects of such a revaluation were not known beforehand. Some assumed that the announcement would have a large impact on other Asian currencies. Others argued that since the news had been expected for some time, all possible effects on the Asian currencies would have already been priced into the market.

\begin{figure*}[tp]
\begin{center}
\includegraphics[width=0.85\textwidth]{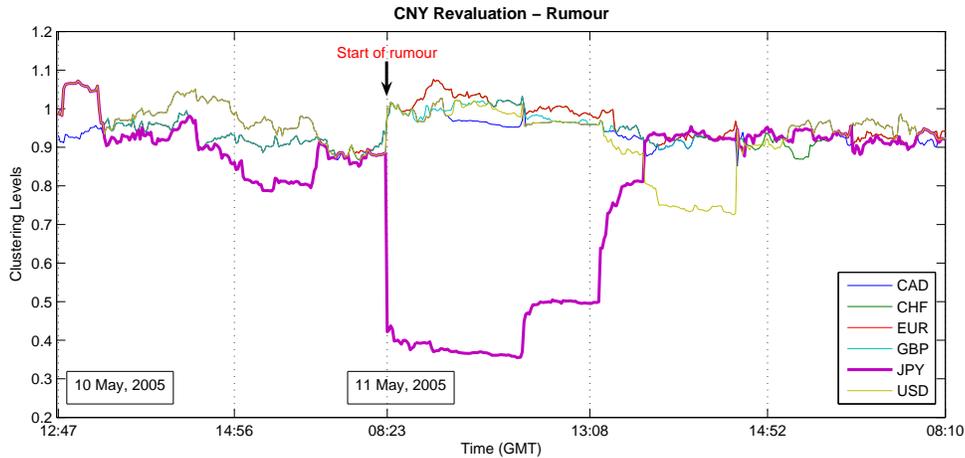}
\end{center}
\caption{(Color online) The clustering distances for the day before, and the day of, the false rumour concerning CNY revaluation.}
\label{fig:Rumour}
\end{figure*}

Figure \ref{fig:Rumour} shows the clustering levels for the day before, and the day of, this false rumour (10--11 May 2005). It is evident that there is a strong dip in the clustering distance for JPY. There are two important points to note. First, the effect is seen in the JPY cluster level from the very first window which incorporates the first 
minute of the rumour. In other words, the market collectively converged to a concensus concerning which currency should be traded -- despite the fact that there was no invisible hand or central controller to conduct the coordination process. Second, the observed clustering lasts for approximately 30 minutes longer than the width of one window, indicating that strong clustering is seen for the entire time that the rumour is prevalent in the market. Note that for such a strong clustering effect to be seen as soon as the window includes the first timestep from the rumour, the returns must be very large in order to induce such a rapid change in the correlation.

\begin{figure}[tp]
\begin{center}
\includegraphics[width=0.45\textwidth]{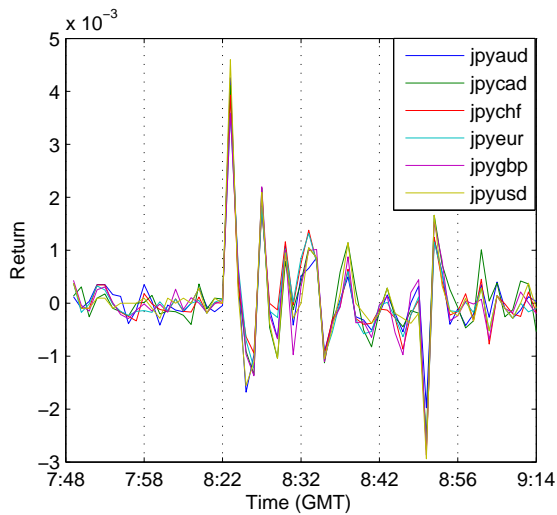}
\end{center}
\caption{(Color online) The JPY-based exchange rate returns for the time including the false CNY revaluation rumour. In the inset, jpyaud denotes the exchange rate between JPY and AUD etc.}
\label{fig:RumourReturns}
\end{figure}

Figure \ref{fig:RumourReturns} shows the returns for the JPY-based exchange rates for the period of time which includes the false CNY revaluation rumour. The degree of synchronization between the different rates once the rumour emerges, is remarkably strong -- even more so than for 11 September. The buying and selling of JPY is dominating any activity in the rest of the currencies included in the analysis.

\subsection{3. CNY Revaluation --- Actual Event}
\begin{figure*}[tp]
\begin{center}
\includegraphics[width=0.85\textwidth]{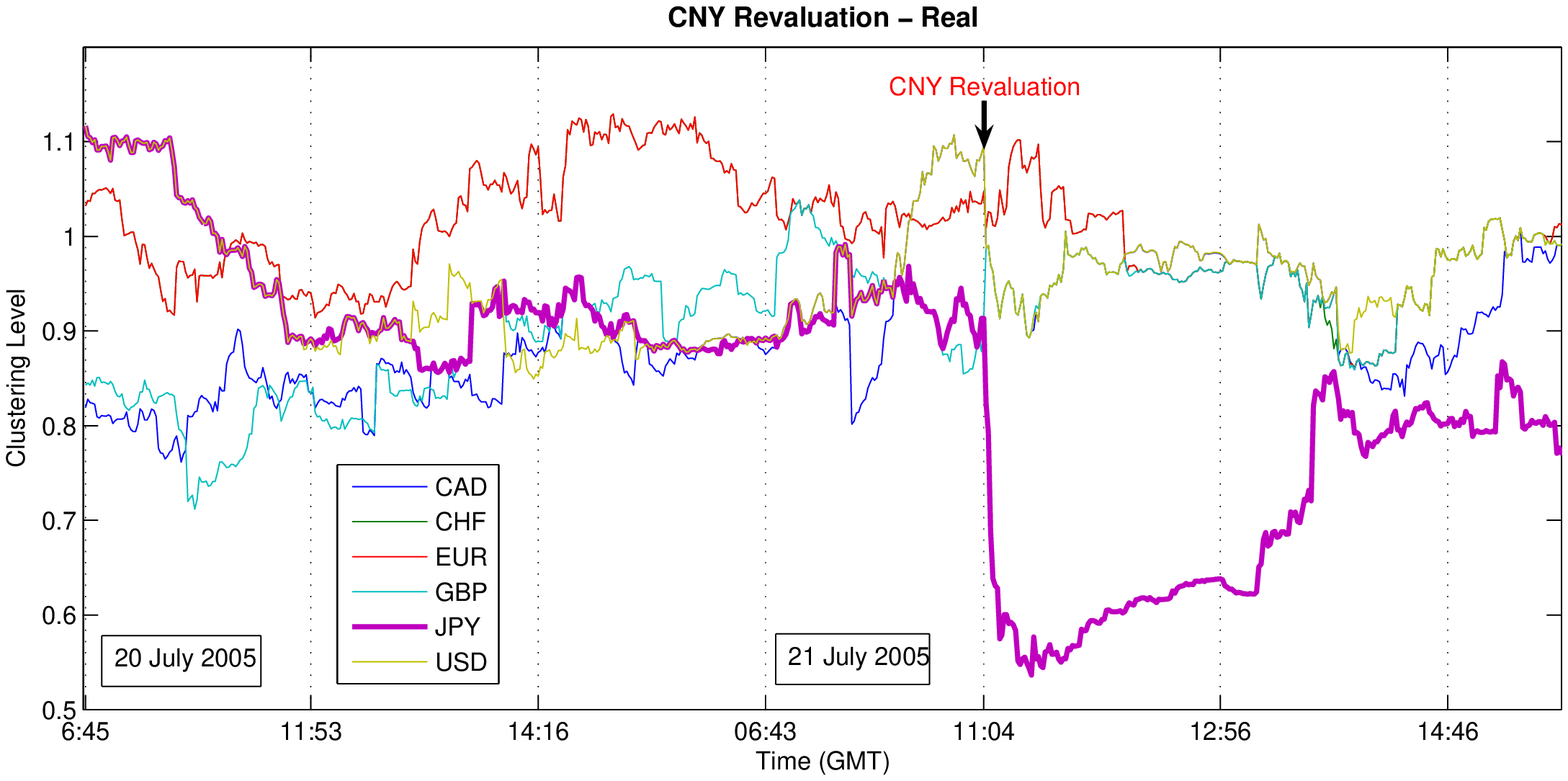}
\end{center}
\caption{(Color online) The clustering distances for the day before, and the day of, the real CNY revaluation.}
\label{fig:Revaluation}
\end{figure*}

Figure \ref{fig:Revaluation} shows the clustering distance for the day before, and the day of, the real CNY revaluation (20--21 July 2005). The similarities with Figure \ref{fig:Rumour} for the rumour, are striking. The market appears to be noisier before the release of the news, which is probably a result of some failed tube bombings in London earlier that same morning. However, the clustering level seen in the JPY is very similar in both the rumour and the real revaluation. Again, the clustering is seen to last for longer than one window length, indicating genuine clustering and not simply one large outlying point. There is also some evidence that the FX market has collectively `learnt' from the case of the rumour, or equivalently that it has some kind of `memory' of what happened before: in particular, Figure \ref{fig:Rumour} shows that the clustering distance for JPY returns to its original level once the rumour becomes discredited, whereas Figure \ref{fig:Revaluation} shows that the clustering distance for JPY remains lower for a long time afterwards. In other words, similar patterns are observed for the rumour and real cases, but the real case evolves to a modified state.

\begin{figure}[tp]
\begin{center}
\includegraphics[width=0.45\textwidth]{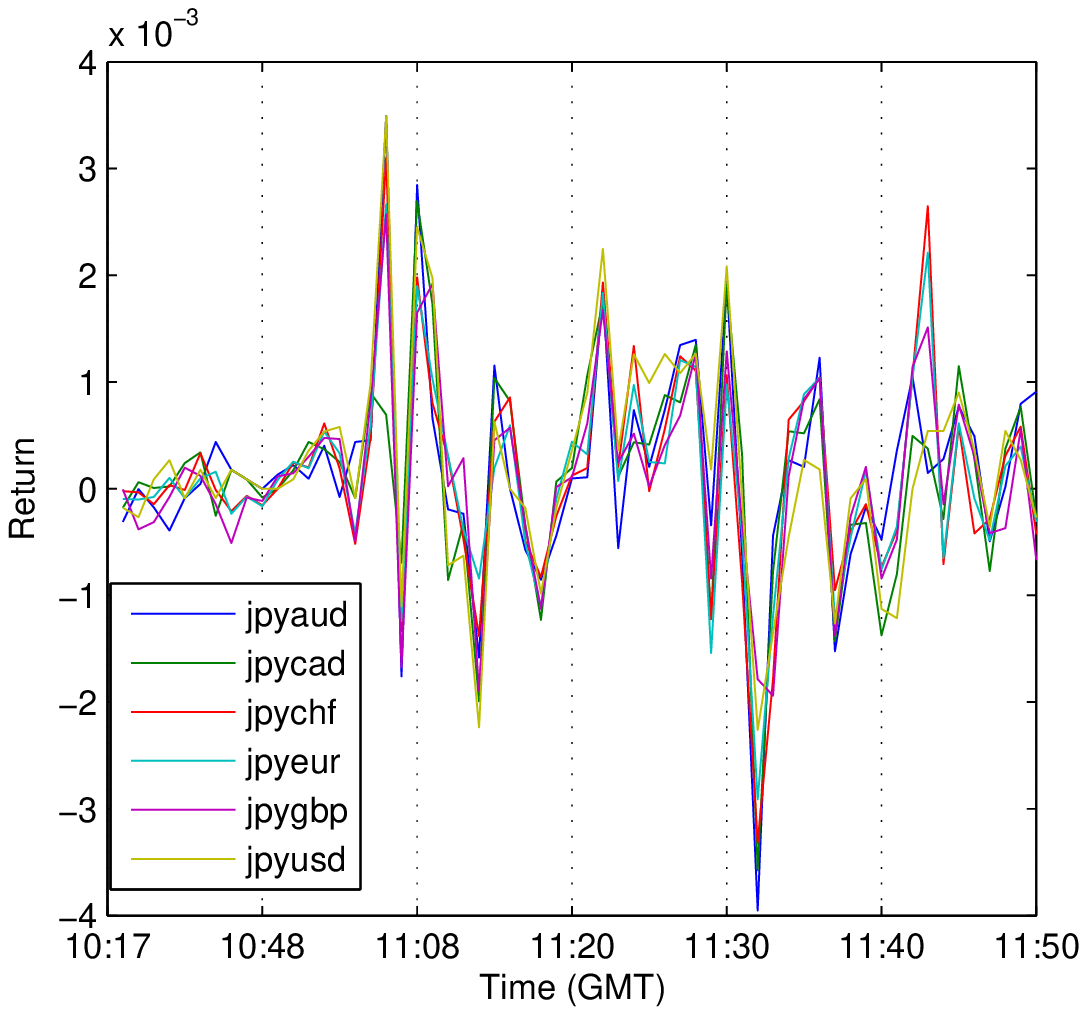}
\end{center}
\caption{(Color online) The JPY-based exchange rate returns for the time period which includes the real CNY revaluation.}
\label{fig:RevalReturns}
\end{figure}

Figure \ref{fig:RevalReturns} shows the JPY-based exchange rate returns for the time period which includes the real CNY revaluation. As with the false rumour, the extent to which the JPY-based exchange rates become synchronized once the news is released, is striking.

\subsection{4. Surprising Economic News}
\begin{figure}[tp]
\begin{center}
\includegraphics[width=0.45\textwidth]{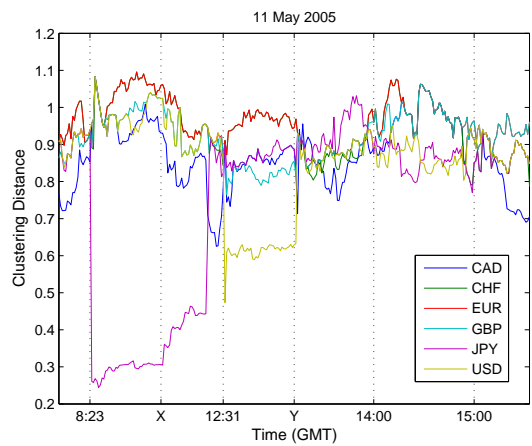}
\end{center}
\caption{(Color online) The clustering distances from using the shortest possible window size (42 returns) for the day of 11 May 2005.}
\label{fig:Rumour42Window}
\end{figure}

Looking again at Figure \ref{fig:Rumour}, and in particular the behaviour later in the day of 11 May, there is a feature which at first seems puzzling. There is a dip in the USD cluster distance which lasts for a short time (it lies within the interval 13:08 to 14:52). This can be understood more easily by referring to Figure \ref{fig:Rumour42Window}, in which we show the clustering distance results from 11 May 2005 using the smallest possible window size ($T$=$N$=42). The time marked `X' is the last time for which the window includes the return at 08:23 GMT. The clustering in JPY continues beyond this time, implying that the JPY cluster is the result of a systematic clustering of JPY-based exchange rates. The time marked `Y' is the last time for which the window contains the result at 12:31 GMT (which is where the USD clustering starts). It can be seen that as soon as the window does not include the timestep at 12:31, there is no clustering. Hence the `clustering' of USD in this situation is simply caused by the behavior at one timestep. But what is this behavior due to?

\begin{figure}[tp]
\begin{center}
\includegraphics[width=0.45\textwidth]{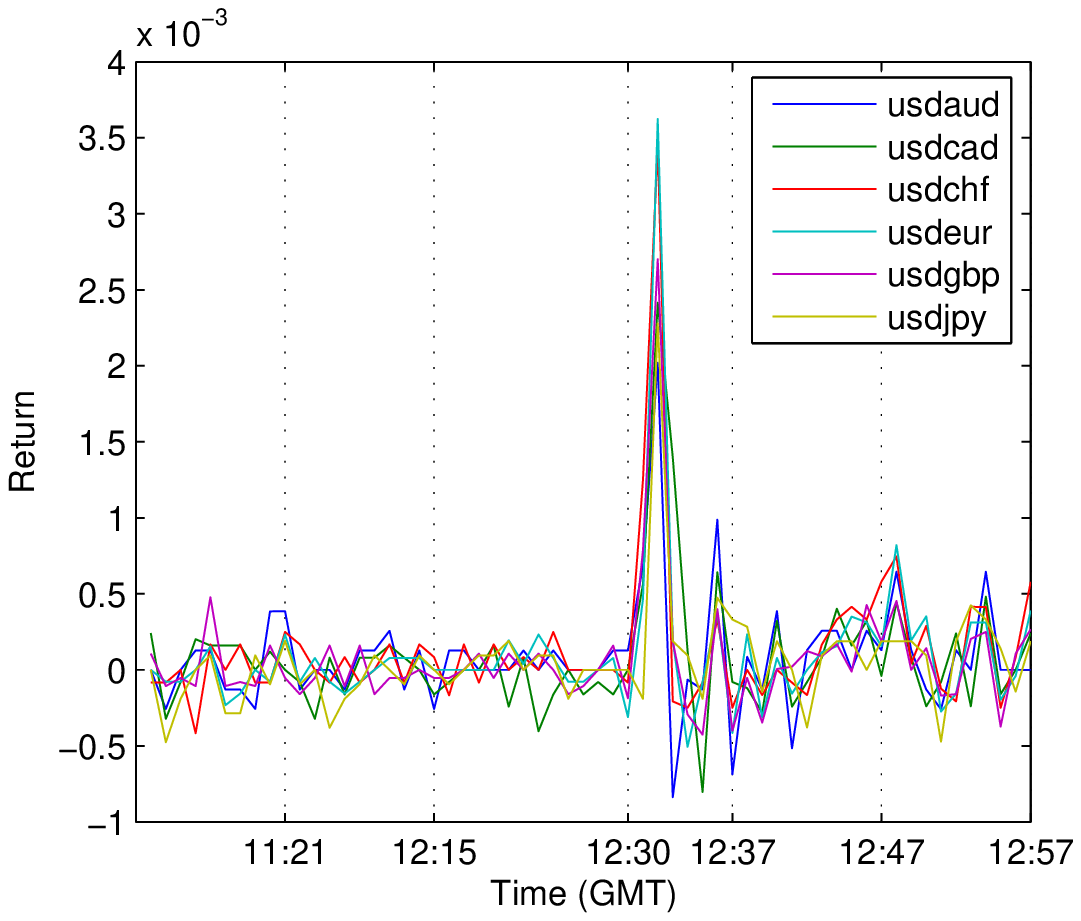}
\end{center}
\caption{(Color online) The returns for the USD-based exchange rates for a time period including the Trade Balance news release on 12 May 2005.}
\label{fig:USDSurpriseReturns}
\end{figure}

It turns out that at 12:30 GMT on 11 May 2005, the Census Bureau of the U.S. Department of Commerce made a very surprising announcement about the U.S. Trade Balance. This caused a very rapid change in the value of USD against all other currencies. The returns for the USD-based exchange rates around the time of the announcement are shown in Figure \ref{fig:USDSurpriseReturns}, showing that between 12:30 and 12:31 GMT, the value of the USD has fluctuated against all the other currencies shown in the analysis \cite{RateDef}. The directional price change for all these exchange rates occurred in less than one minute, whereas it appears that the volatility of the exchange rates is larger following the announcement. This is in agreement with papers researching the impact of economic announcements on financial markets \cite{EL1993}.

\section{V. Robustness of Results}

The results detailed above are all for the SLCA/MST clustering method. As we mentioned, there are potential problems with using this method if the data structure is not appropriate. In this section we repeat the analysis for three different clustering algorithms: ALCA, CLCA and Ward. Note that the Ward algorithm is constrained to use the Euclidean distances between timeseries, and not the standardized Euclidean distance. Thus the distances seen in the Ward algorithm graphs are not the same distances as in the other graphs.

\begin{figure*}[tp]
\begin{center}
\includegraphics[width=0.95\textwidth]{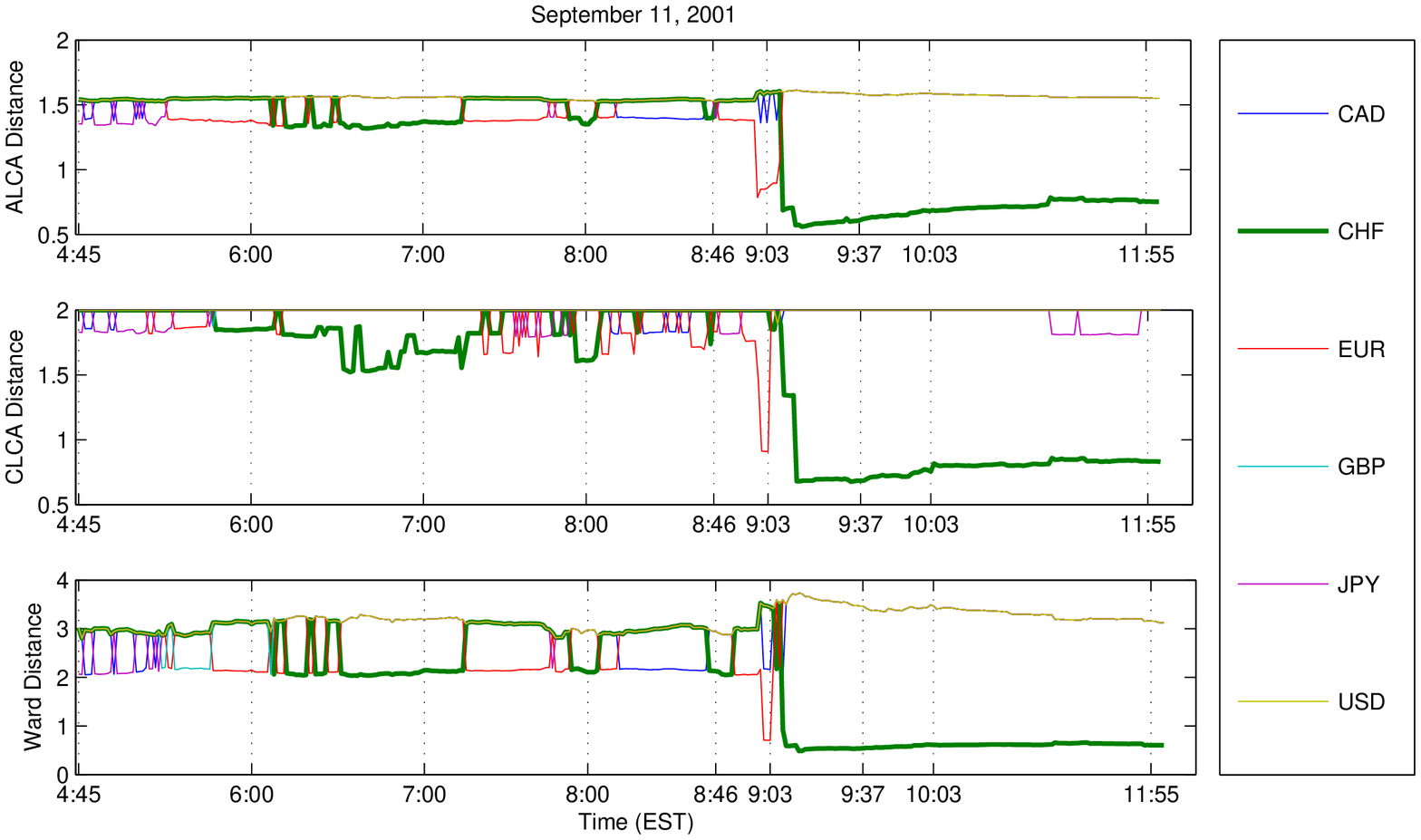}
\end{center}
\caption{(Color online) 11 September cluster distances using three alternative clustering algorithms.}
\label{fig:Sept11MultiPanel}
\end{figure*}

Figure \ref{fig:Sept11MultiPanel} shows the results for 11 September. The top panel is the ALCA distance, the middle panel is the CLCA distance and the bottom panel is the Ward distance. There are several points of interest in this figure. First, the initial clustering of EUR, which then becomes dominated by the CHF cluster, is evident in all three panels. Second, the CLCA distances (middle panel) show that prior to the attacks the cluster distance is consistently close to 2.0 for many currencies, for most of the time, corresponding to perfect anticorrelation and hence $\hat{d}_{ij} = -1.0$. In general this occurs for the returns of two inverse exchange rates (for example USD/CHF and CHF/USD). For the cluster distance for a currency to be 2.0 it means that by the time all the exchange rates with that currency as the base are in the same cluster, there are two inverse exchange rates in the cluster. It is obvious that one cannot consider two inverse exchange rates to move together, so this level of clustering must be interpreted as zero level of clustering. This allows us to distinguish real clustering from noise.

\begin{figure*}[tp]
\begin{center}
\includegraphics[width=0.95\textwidth]{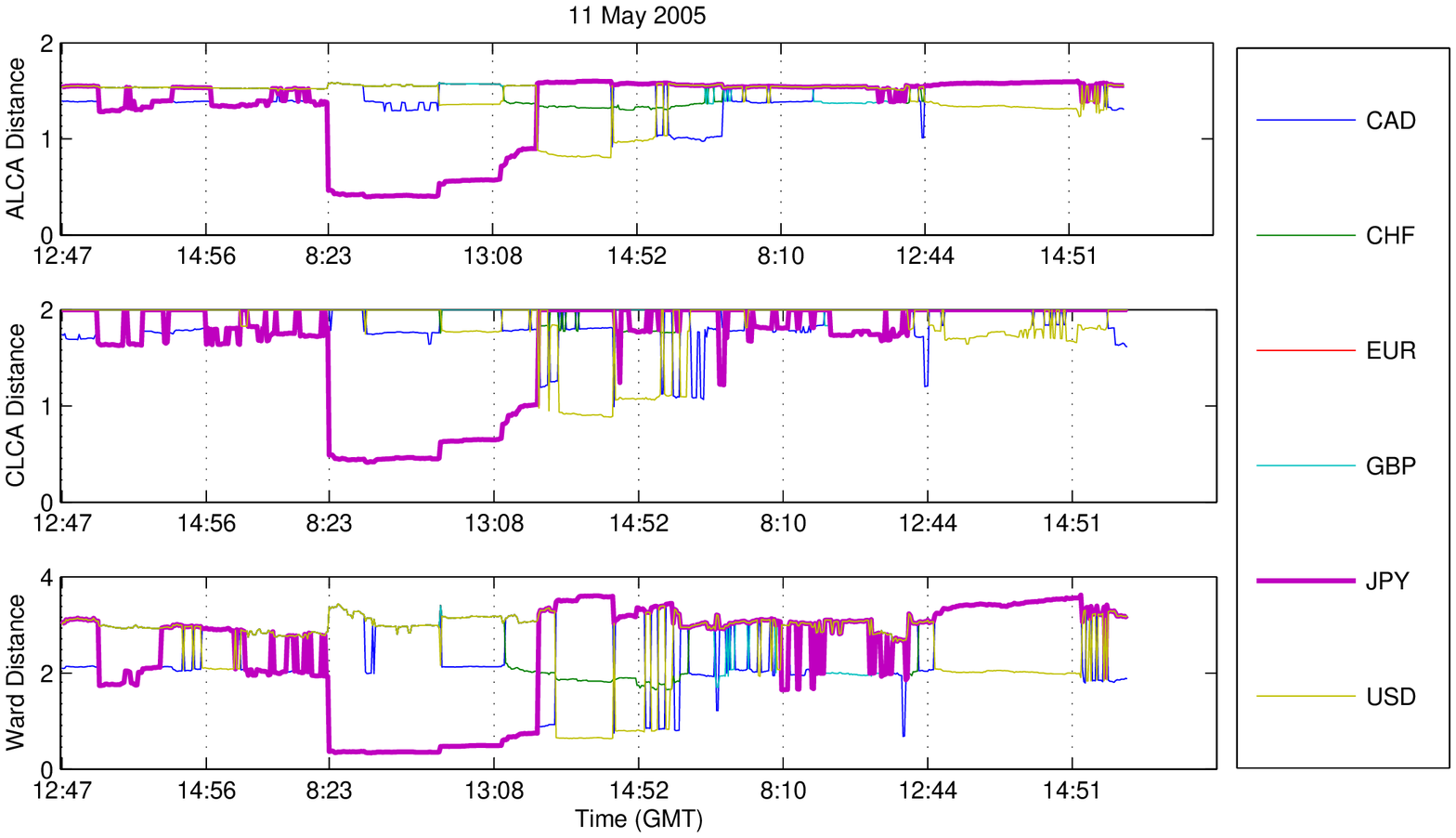}
\end{center}
\caption{(Color online) The clustering distances for the days of 10--11 May 2005, using three alternative clustering algorithms.}
\label{fig:RumourMultiPanel}
\end{figure*}

\begin{figure*}[tp]
\begin{center}
\includegraphics[width=0.95\textwidth]{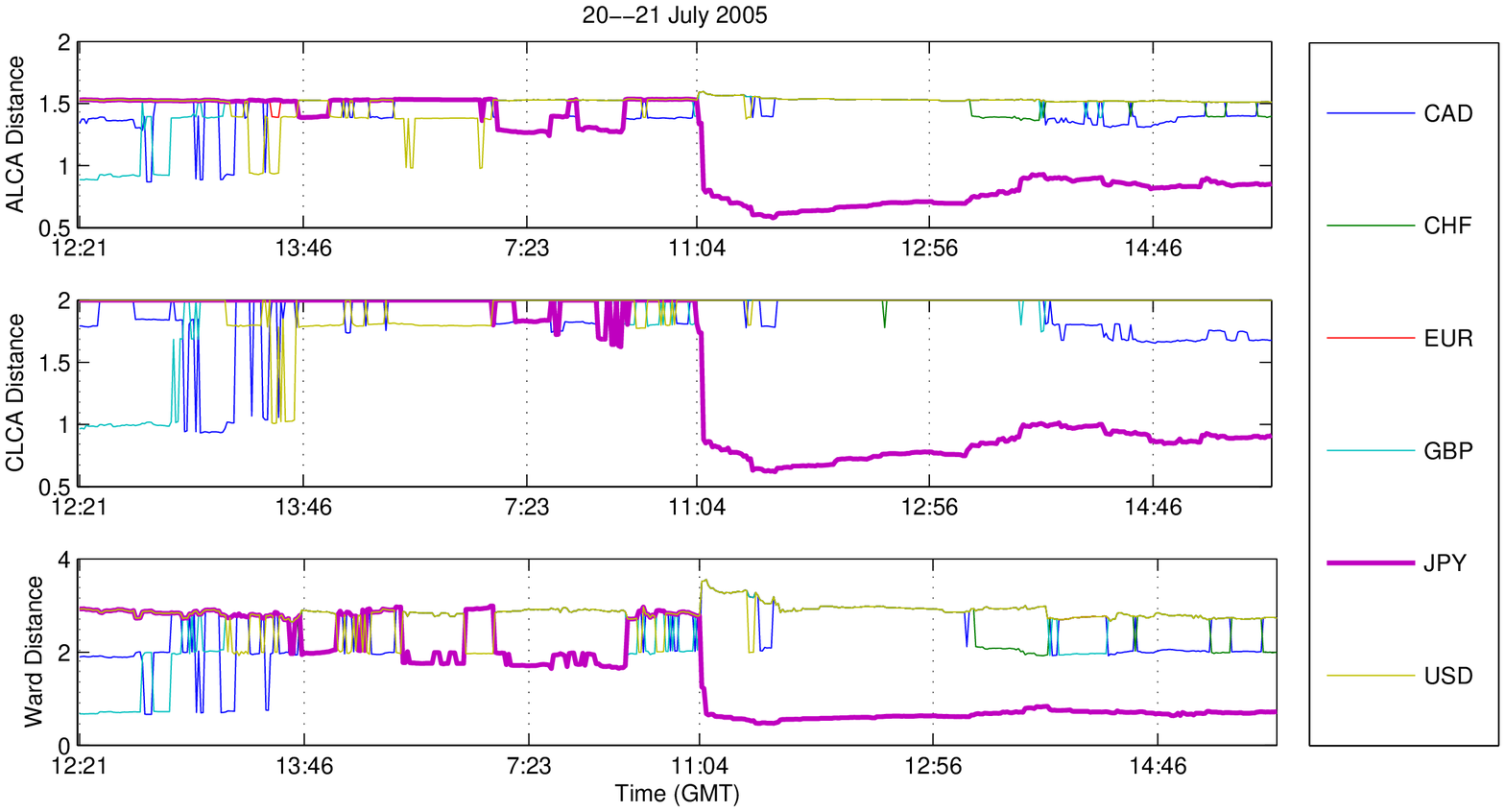}
\end{center}
\caption{(Color online) The clustering distances for the days of 20--21 May 2005, using three alternative clustering algorithms.}
\label{fig:RealMultiPanel}
\end{figure*}

Figures \ref{fig:RumourMultiPanel} and \ref{fig:RealMultiPanel} shows the equivalent figures for the case of the false CNY revaluation rumour and the actual revaluation respectively. Similar effects can be seen in these figures also. It can be seen that the clustering observed as a result of the news can be distinguished from the case of zero clustering. In addition, it can be seen that the clustering levels from the CLCA and Ward algorithms are quite noisy, but still show the same clustering for JPY.

\section{VI. Toward a Taxonomy of News}
One very practical outcome of this research, would be to understand what types of news move markets and in what types of ways. In particular, we would like to classify news in terms of its different types of impact. Based on the observed synchronization and oscillatory ripple-like effects in Sec. IV, we can hope to develop general shock-response and vulnerability indicators in response to the different types of events which a particular news agency could report. We are still short of this goal. Indeed we would need to expand the database of case-studies quite considerably in order to pin down such classifications. However if we allow ourselves to speculate for a moment, guided by the four case-studies in Sec. IV (and some subsequent studies which we have carried out but which are not presented in the current paper), we would suggest that news could be usefully classified as belonging to one of the following possible `species' \cite{species}:

\noindent {\em Species A}: News that is (1) unexpected in terms of the fact that it occurs at all, (2) surprising in the sense of when it occurs, (3) is not related directly to markets, (4) has never happened before, and (5) is true. The events of 11 September fall into this category.  

\noindent {\em Species B}:
News that is (1) somewhat expected in terms of the fact that it occurs at all, (2) surprising in the sense of precisely when it occurs, (3) is not related directly to markets, (4) has never happened before, and (5) is true. It turns out that the 7/7 terrorists attacks in London, fall into this category. The fact that the U.K. was a close ally of the U.S. and had hence received terrorist threats since the Iraq invasion, meant that such attacks might have been expected at some stage.

\noindent {\em Species C}:
News that is (1) somewhat expected in terms of the fact that it occurs at all, (2) surprising in the sense of precisely when it occurs, (3) is related directly to markets, (4) has never happened before, and (5) is untrue. The CNY rumour falls into this category.
Although such a revaluation had been expected for a while, the timing of any official announcement was completely unknown to the markets in advance -- hence such a rumour would have initially seemed highly credible to traders.

\noindent {\em Species D}:
News that is (1) somewhat expected in terms of the fact that it occurs at all, (2) surprising in the sense of precisely when it occurs, (3) is related directly to markets, (4) has happened before, and (5) is true. The real CNY revaluation falls into this category. 

Our results to date suggest that markets tend to respond in a remarkably similar way to particular species of news. In the case of terrorist attacks, for example, the global response to the London attacks represented a much milder but similar response to the earlier U.S. attacks. For the case of the rumour and real revaluation of the CNY, a similar pattern of response was observed. This would not be a surprise for a mechanical object such as a spring-balance: a spring-balance will respond in the same way if it is pushed down in the same way. For a spring-balance this isn't surprising, since nothing much else is happening to it. But for this to happen in a market, is extraordinary. Indeed it represents a significant endorsement of the whole idea that a financial market can be seen as a real entity -- a Complex System in its own right. 

\section{VII. Conclusion}
To summarize, we have shown that the clustering distance is a useful tool for investigating the collective human response to news in the FX market. In particular we have shown that there is a markedly different response seen when the news is genuinely unexpected, as opposed to a scheduled economic announcement that has a value far from what was expected by the market. We believe that this tool could eventually prove very useful for classifying news events.

We have also confirmed the robustness of the results from the MST method applied to the FX data. This is particularly useful since it justifies the use of the intuitive graphical representation provided by the MST. Indeed, the clustering distances from the MST are less noisy than for the other methods investigated. Section V illustrates the importance of comparing results with alternative clustering methods, so as to be able to distinguish real clustering from noise.


\begin{thebibliography}{99}
\bibitem{ISOCode}
These currency codes are standard in the financial industry. They have been established by the International Organization for Standardization (ISO) and are known as ISO 4217 \cite{ISO4217}.
\bibitem{inverseCluster}
If one looks closely, one can see the equivalent cluster of X/CHF rates in the hierarchy.
\bibitem{RateDef}
Recall that the exchange rate USD/X is the number of units of X that one USD can buy. Hence, if USD/X decreases, USD is less valuable with respect to currency X.
\bibitem{species} This list of different `species' of news is not meant to be exhaustive, but covers many types of news that are of major interest.
\end{thebibliography}
\end{document}